\documentclass[useAMS]{tMPH2e}
\usepackage{rotating}

\begin{document}
\doi{10.1080/00268970701466261}
\issn{}
\issnp{} \jvol{105} \jnum{11-12} \jyear{2007} \jmonth{10 June - 20 June}
\setcounter{page}{1723}
\markboth{N. Vanhaecke and O. Dulieu}{Precision measurements with polar molecules...}

\title{{\itshape 
Precision measurements with polar molecules: the role of the black body radiation
}}

\author{N.VANHAECKE$^{\ast}$\thanks{$^\ast$Corresponding author. Email:
nicolas.vanhaecke@lac.u-psud.fr} \vspace{6pt} and O. DULIEU \\
\vspace{6pt}  Laboratoire Aim\'e Cotton, CNRS, B\^at. 505, Campus d'Orsay, 91405 Orsay Cedex,
France\\\vspace{6pt}\received{Received 28 February 2007; in final form 11 May 2007} }

\maketitle

\begin{abstract}
In the perspective of the outstanding developments of high-precision measurements of fundamental constants using polar molecules related to ultimate checks of fundamental theories, we investigate the possibly counterproductive role of the black body radiation on a series of diatomic molecules which would be trapped and observed for long durations. We show that the absorption of the black body radiation at room temperature may indeed limit the lifetime of trapped molecules prepared in a well-defined quantum state. Several examples are treated, corrresponding to pure rotational absorption, pure vibrational absorption or both.
We also investigate the role of the black body radiation induced energy shift on molecular levels and how it could affect high-precision frequency measurements.

\bigskip

\end{abstract}

\section{Introduction}
Over the last decades more and more efforts have been put on searching experimentally for "new physics"
breaking the fundamental $C$,$P$,$T$ symmetries.
High-energy experiments have initiated this quest for physics beyond the Standard Model more than
fourty years ago with the discovery of the $CP$-violation \cite{Christenson1964}.
Since then not only high-energy experiments have been searching for parity violations, but also
low-energy experiments based on high-precision measurements, as initially proposed in atomic systems
\cite{Bouchiat1975}.
For instance, recent results on the $P$ parity violation have been obtained probing the highly
dipole-forbidden 6$S$-7$S$ transition in cesium \cite{Guena2005,Bennett1999}.

This kind of precision measurement is especially eager to take advantage of the most sensitive tools to
gain in precision and to reduce systematic errors. 
In that respect, cold atoms have brought huge improvements in several high-precision investigations,
and nowadays cold molecules hold even greater promise.
In such experiments, interferometers and quantum beats with atoms and molecules are often used and very
long interrogation times (up to a few seconds) are needed to let the phase being accumulated in the interferometer.
The search for time dependence of the fundamental constants needs particularly high-precision
measurements since it is looking for small relative drifts.
In the search for a time dependence of the fine structure constant, which would imply the violation of both
Lorentz invariance and {\it CPT} symmetry, the best measurement to date has been achieved through
comparisons of atomic frequency standards \cite{Peik2004}.
The recent high-resolution spectroscopy of cold OH radicals in a Stark-decelerated beam combined with
astrophysical measurements of OH megamasers gives hope to yield an even better determination of the
time variation of the fine structure constant \cite{Hudson2006}.
The time dependence of the proton-electron mass ratio is planned to be investigated by
measuring extremely precisely the inversion splitting in the ammonia molecule in a molecular fountain \cite{priv:Rick}.
A tremendously exciting search for a possible permanent electric dipole moment (EDM) of atoms, nucleons
and elementary particles like electron and muon is being carried out
\cite{Romalis2001,Guest2007,Baker2006,Regan2002,Farley2004}, as its existence would reveal a violation
of the time-reversal invariance.
To date, the lowest limit on the value of a possible electron EDM has been established using the
thallium atom \cite{Regan2002}.
Molecules are expected to open the way to even more precise measurements and many
efforts are put nowadays to set up experiments with cold beams of heavy polar molecules such
as YbF, PbO or PbF \cite{Hudson2002,DeMille2000}. 
Cold slowed or trapped molecules offer even better prospects, allowing very long interrogation times on
the order of several seconds.

The accuracy of such high-precision measurements has become so amazingly high, that every possible little 
disturbance has to be considered very carefully.
So has to be the black body radiation (BBR), always present in an experimental environment.
Atoms in the ground state are normally immune to the absorption of black body photons at room temperature, since such
photons do not carry enough energy to induce a dipole-allowed transition. 
This is not the case in Rydberg species, as it is known for a long time \cite{Gallagher1979}.
A distribution of molecular ions has also shown to evolve toward a thermal equilibrium at the environment
temperature through exchange of black body photons \cite{Hechtfischer1998}.
It has been proposed to use BBR to cool down internal states of molecular ions \cite{Vogelius2002} and
very recently laser-trapping of radium took advantage of the BBR as repumping light \cite{Guest2007}.
Recently BBR absorption rates have been experimentally measured in trapped clouds of cold OH and OD radicals \cite{Hoekstra2007}.
The presence of the BBR induces also a shift of the atomic resonances, which is a well-known effect on frequency measurements in atomic clocks.
Such a light shift is now one of the biggest uncertainties in high-precision measurements of frequencies \cite{Angstmann2006a,Beloy2006,Porsev2006}.
So far it has not been observed in molecular systems.

In this paper we investigate the possible limitations induced by the BBR on experiments aiming at interrogating molecular systems over long time intervals. 
Such experiments often require heavy molecules with large permanent dipole moments which make them very sensitive to the presence of any external electric field, like the BBR electric field. 
We first recall in section \ref{sec:model} the different transitions which can be induced by BBR in molecules, emphasizing through a simple model the crucial role of both the {\it magnitude} and the {\it variation} of the permanent electric dipole moment of the molecule with the internuclear distance. 
Then we explore several classes of molecules of relevance for running, planned or possible high-precision measurements: heavy fluoride radicals (section \ref{sec:ionic}),
strongly polar alkali dimers, alkali hydrides as well as sulfur oxide (section \ref{sec:other}). 
We investigate in section \ref{section:lightshift} the role of the BBR-induced energy shift in two molecular systems of relevance now in the context of high-precision measurements: the YbF radical and the ammonia molecule.

\section{Absorption of black body radiation by diatomic molecules}
\label{sec:model}

\begin{figure}
\begin{center}
\epsfbox{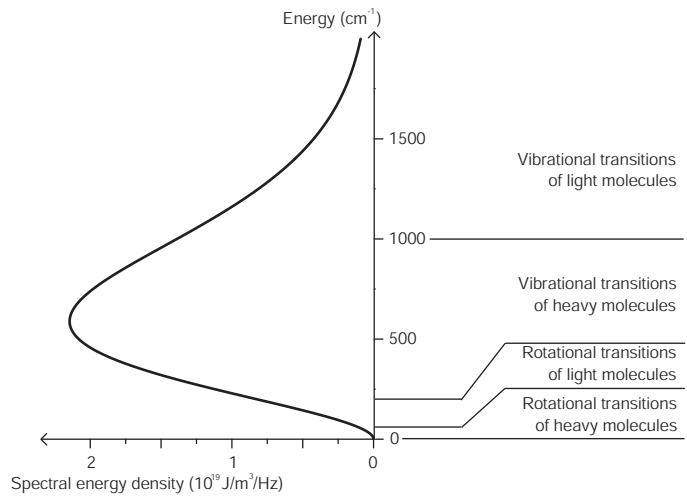} 
\caption{Spectral energy density of the BBR at room temperature (300\,K). On the right side are schematically depicted the typical energy ranges for vibrational and rotational transitions in heavy and light molecules. }
  \label{fig:bbr}
\end{center}
\end{figure}

The spectral energy density of the BBR at a frequency $\nu$ and a temperature $T$ is given the Planck
radiation law:

\begin{equation}
\rho(\nu) 
= 
\frac{8\pi h \nu^3}{c^3}
\frac{1}{e^{h \nu / k_B T}-1}.
\end{equation}

Figure \ref{fig:bbr} shows the BBR spectral energy density at room temperature.
The BBR absorption rate of a quantum system in the initial state $i$ reads:

\begin{equation}
\Gamma_i^\text{BBR abs} 
= \sum_{f}\frac{\mu_{fi}^2}{6\epsilon_0 \hbar^2}\rho(\nu_{fi}) 
\equiv \frac{8 \pi^2}{3 \epsilon_0 \hbar c^3}
\sum_{f}
\frac{\nu_{fi}^3 \mu_{fi}^2}{e^{h \nu_{fi} / k_B T}-1},
\label{eq:rate}
\end{equation}

where the summation is performed over all final states $f$ accessible from the initial state $i$ {\it
via} an electric dipole transition at frequency $\nu_{fi}$. The details of the rovibronic
structure of the quantum system are involved through the matrix elements $\mu_{fi}=\big< f \big|
\mu \big| i\big>$ of the dipole moment operator $\mu$ between the initial
state $i$ and the final state $f$. We restrict ourselves to the study of heteronuclear (i.e. polar)
diatomic molecules, assuming that both $i$ and $f$ states can properly be described within the
Born-Oppenheimer (BO) approximation with total wave functions $\Psi^{BO}_i$ and $\Psi^{BO}_f$. The BO
wave functions are written as products of electronic ($\psi$), vibrational ($\chi$), and rotational
wave functions ($\Theta$):

\begin{equation}
\Psi^{BO}_{i,f} = \psi_{\alpha_{i,f}}(R;\left\{\vec{r}\right\}) \chi_{v_{i,f}}(R)
\Theta_{J_{i,f}}(\Omega) .
\label{eq:BO}
\end{equation}

In eq.(\ref{eq:BO}), the $\alpha_{i,f}$, $v_{i,f}$ and $J_{i,f}$ labels refer respectively to the
electronic, vibrational, and rotational quantum numbers of the initial and final states. The electronic
part depends on all electronic coordinates $\left\{\vec{r}\right\}$, and the rotational part on the
angular coordinates $\Omega$ of the molecular axis in the space-fixed frame. 
We assumed that there is no coupling between vibrational and
rotational degrees of freedom.
The transition dipole moment
$\mu_{fi}$ can be reformulated as the integral over the interatomic separation $\vec{R}$: 

\begin{equation}
\mu_{fi} = \int_{}^{}{\chi_{v_{f}}(R) \Theta_{J_{f}}(\Omega)
 \vec{\mu}^{BO}_{fi}(\vec{R}) \cdot \vec{\epsilon} \;\chi_{v_{i}}(R) \Theta_{J_{i}}(\Omega)dR d\Omega},
\label{eq:mu}
\end{equation}

which accounts for the orientation of the instantaneous dipole moment for all electrons relative to
the polarization $\vec{\epsilon}$ of the black body photon. The $R$-dependent transition dipole
moment function $\vec{\mu}^{BO}_{fi}(\vec{R})$ is obtained after integration of the product
$\psi_{\alpha_{f}}\mu\psi_{\alpha_{i}}$ on the electronic coordinates.

The energy distribution of the BBR spectrum peaks around 600\,cm$^{-1}$ at room temperature,
which is not sufficient to induce any electronic transition starting from the ground state or from a
low metastable state in diatomic molecules. Therefore we can safely ignore them, and restrict our
attention to transitions between rotational or vibrational levels of a single electronic state
$\alpha$. The transition dipole moment function now reduces to the permanent dipole moment function
$\mu_{\alpha}(R)$ of the state $\alpha$ so that eq.(\ref{eq:mu}) becomes

\begin{equation}
\mu_{v'J' vJ} \propto \int_{0}^{\infty}{\chi_{v'}(R) \mu_{\alpha}(R) \chi_{v}(R) dR}
\equiv \mu_{v' v},
\label{eq:mu_v}
\end{equation}

where $v$ and $v'$ (resp. $J$ and $J'$) are the initial and final vibrational (resp. rotational) levels
of the transition within the state $\alpha$. The proportionality factor in eq.(\ref{eq:mu_v}) is the
H\"{o}nl-London factor between the initial and the final state, which accounts for the initial and
final rotational levels $J$ and $J'$ and for the relative orientation of the molecular axis and of the
polarization of the BBR electric field. 
We will see from
eq.(\ref{eq:mu_v}) that depending on the type of transition, the magnitude of $\mu_{v'J' vJ}$ is
determined either by the magnitude or by the $R$-dependence of the dipole moment function. The knowledge of molecular
dipole moment functions generally require quantum chemistry calculations which are available over a wide
$R$-range only for a few molecules of interest here. The value of this function at the equilibrium
distance $R_e$ of the molecule and its first derivative around $R_e$ are also sometimes given in the
literature. This information is actually sufficient to evaluate the BBR absorption rate, as in all
cases considered here the vibrational ground state wave function -which peaks at $R_e$- is of relevance.
Indeed, the function $\mu_{\alpha}(R)$ can be expanded to the first order around $R_e$ and reads

\begin{equation}
\mu_{\alpha}(R) = \mu_{\alpha}(R_e) + \Big[\frac{d \mu_{\alpha}}{ d R}\Big]_{R=R_e}(R-R_e) \, , 
\label{eq:taylor}
\end{equation}

where $\Big[\frac{d \mu_{\alpha}}{ d R}\Big]_{R=R_e}$ is the first derivative of the electric dipole
moment function evaluated at $R_e$. The scheme displayed in Fig.\ref{fig:bbr} illustrates the various
situations treated here. For light molecules, with a mass smaller than ten atomic mass units say, the
typical energy for vibrational transitions often exceeds the peak of the BBR energy distribution,
so that pure rotational transitions are dominant, the molecule remaining in its
lowest vibrational level ($v=0$). The constant term $\mu_{\alpha}(R_e)$ in eq.(\ref{eq:taylor}), i.e. the
magnitude of the dipole moment, brings the main contribution to $\mu_{v J' vJ}$ and the integral in
eq.(\ref{eq:mu_v}) almost reduces to $\mu_{\alpha}(R_e)$. In contrast, for heavier molecules the
energy of rotational transitions corresponds to the low energy part of the BBR distribution, so that
vibrational transitions are favoured.
Therefore $\mu_{v'J' vJ}$ (with $v'\neq v$) mainly depends on the slope of the dipole moment function, and no more on its
magnitude, due to the mutual orthogonality of vibrational wave functions. A further simplification can be made
if the vibrational (harmonic) constant $\omega_e$ of the relevant potential curve is known. The radial
integral between two adjacent vibrational levels  in eq.(\ref{eq:mu_v}) can be analytically evaluated for
a harmonic potential:

\begin{equation} 
\mu_{v+1, v} = \sqrt{\frac{v+1}{2}} \sqrt{\frac{\hbar}{m \omega_e}}
\Big[\frac{d \mu}{ d R}\Big]_{R=R_e},
\label{eq:vvp1}
\end{equation}

where $m$ is the reduced mass of the molecule. These approximations break down if the potential
well is strongly anharmonic, if the dipole moment function is notably nonlinear or too flat around
$R_e$, or if couplings such as rovibrational couplings are not negligeable.

\section{BBR absorption rate for fluoride radicals: BaF, YbF, HgF and PbF}
\label{sec:ionic}

In heavy paramagnetic molecules such as BaF, YbF, HgF and PbF - of relevance for the search of the
electron EDM - the vibrational wave function of low-lying vibrational levels extends typically over a
small fraction of an atomic unit (1\,a.u.$\equiv a_0=0.0529177$\,nm), so that the dipole moment function
of the electronic ground state can safely be considered linear over this extension. 
Therefore, the strength of the electric dipole transition between two consecutive vibrational states
can be estimated following eq.(\ref{eq:vvp1}) as soon as the derivative of the electric dipole moment
function at the equilibrium distance is known. 

Diatomic lanthanide compounds possess partially filled 4$f$ and 5$d$ shells and involve strong
inter-shell couplings, which generate very complicated spectra. Their study is therefore challenging
for both theoretical and experimental investigations.
Among the molecules mentioned above, YbF \cite{Dolg1992,Liu1998} and PbF \cite{Das2002} have been
studied from {\it ab initio} theory and dipole moment function of YbF and its value and first
derivative for PbF at the equilibrium internuclear distance have been derived. 

Extensive quantum chemistry studies on alkaline earth monohalides are available mainly for the lighter ones (involving Be, Mg, Ca, Sr linked with F or Cl), providing all necessary quantities for the present purpose, including the derivative of the permanent dipole moment  \cite{Langhoff1986}. Quantum chemistry calculations have also been devoted to BaF \cite{Arif1996} and HgF \cite{Dmitriev1992}
molecules, providing the main properties of the electronic ground state except the $R$-variation of its
permanent dipole moment function. Fortunately, alkaline earth monohalides (hereafter
labelled as Me-Hal) have been extensively studied in the past, since their theoretical description is
underlined by a simple picture: such molecules are ionic-bond molecules, in which the lone electron
evolves in the field of two atomic ions, namely the closed-shell doubly-charged metal cation Me$^{++}$
and the halogen anion Hal$^-$. The ligand approach \cite{Rice1985,allouche1993} assumes that the electronic
structure of the system can be described with orbitals centered on the free Me$^{++}$ ion while the
electrostatic field of the ligand Hal$^-$ ion is treated as a perturbation. The electrostatic model
proposed by Rittner \cite{Rittner1951} accounts for the mutual polarization of the Me$^{+}$ and Hal$^-$
ions through induced dipole moments which reduce the primary permanent dipole moment of the molecule.
However, due to the large polarizability of the Me$^{+}$ ion, higher order effects break this
approximation down. The more elaborated electrostatic model of T\"orring {\it et al} \cite{Torring1984}
accounts for the strong hybridization of the external electron of the  Me$^{+}$ ion induced by Hal$^-$,
which acts as a radial shift of the center of charge of the electron wave function. The induced dipole moments are
modified accordingly in the expression of the net dipole moment of the molecule, whose derivative can
also be derived. Finally, as pointed out in refs.\cite{Torring1984,Arif1996} an effective polarizability of
the Hal$^-$ ion should be introduced in such ionic molecules which accounts for the influence of the Me$^{+}$ ion on the somewhat floppy negative ion. The effective value of 4.7~a.u. is recommended in refs.\cite{Torring1984,Arif1996} for F$^-$ within the CaF and BaF molecules, indeed much smaller than the free ion polarizability (about 16~a.u. \cite{diercksen1982,Kucharski1984}). 

Using the value 81\,a.u. of ref.\cite{Torring1984} for the Ba$^+$ polarizability, the ionic model yields for BaF at the equilibrium distance ($R_e=4.09$\,a.u.) of its $X ^2\Sigma^+$ electronic ground state $\mu(R_e)= 1.38$\,a.u. (close to the experimental value 1.24\,a.u. \cite{Ernst1986}) and $\Big[\frac{d \mu}{ d R}\Big]_{R=R_e} = 1.78$\,a.u.. For HgF in its $X ^2\Sigma^+$ electronic ground state ($R_e =3.87$\,a.u.) the same model leads to $\mu(R_e)=2.85$\,a.u. and $\Big[\frac{d \mu}{ d R}\Big]_{R=R_e} = 1.63$\,a.u., taking the Hg$^+$ polarizability 19\,a.u. from ref.\cite{Angstmann2006}. 

\begin{table}
 \begin{center}
  \tbl{Equilibrium distances, vibrational and rotational frequencies, values of the dipole moment function and of its first derivative at the equilibrium distance for the dimers discussed throughout the paper. The state considered here is always the rovibrational ground state, unless otherwwise stated.}
 {\begin{tabular}{l|l|l|l|l|l|l|l}\toprule
   Species & Initial state 
   & $R_e$ & $\omega_e$ & $B_e$
   & $\mu_e$ & $\mu_e'$ 
   & Ref  \\
   &  
   & ($a_0$) & (cm$^{-1}$) & (cm$^{-1}$)
   & (D) & (D\,$a_0^{-1}$) 
   &  \\
  \toprule
   YbF  & $X^2\Sigma^+$ & 3.80 & 502 & 0.24 & 3.55 & 3.09 & \cite{Dolg1992} \\
   BaF  & $X^2\Sigma^+$ & 4.09 & 469 & 0.21 & 3.51 & 4.52 & \cite{Torring1989} \\
   HgF  & $X^2\Sigma^+$ & 3.87 & 489 &      & 7.24 & 4.14 &  \\
   PbF  & $X^2\Sigma^+$ & 3.94 & 530 &      & 4.32 & 3.38 & \cite{Das2002} \\
  \toprule
   LiRb & $X^1\Sigma^+$ & 6.50 & 185 & & 4.15 & 0.14 & \cite{Aymar2005} \\
   LiCs & $X^1\Sigma^+$ & 6.82 & 164 & & 5.44 & 0.45 & \cite{Aymar2005} \\
   NaRb & $X^1\Sigma^+$ & 6.84 & 107 & & 3.30 & 0.20 & \cite{Aymar2005} \\
   NaCs & $X^1\Sigma^+$ & 7.20 & 98.0& & 4.61 & 0.34 & \cite{Aymar2005} \\
   KRb  & $X^1\Sigma^+$ & 7.64 & 75.5& & 0.61 & 0.03 & \cite{Aymar2005} \\
   KCs  & $X^1\Sigma^+$ & 8.01 & 66.2& & 1.90 & 0.11 & \cite{Aymar2005} \\
   RbCs & $X^1\Sigma^+$ & 8.28 & 49.4& & 1.23 & 0.07 & \cite{Aymar2005} \\
   \toprule
   LiH  & $X^1\Sigma^+$, $J$=1 & 2.98 & 1405 & 7.5 & 5.88 & 0.96 & \cite{Aymar2007}\\
   CsH  & $X^1\Sigma^+$, $J$=1 & 4.50 & 891  & 2.7 & 8.30 & 2.10 & \cite{Aymar2007}\\
   SO   & $X^3\Sigma^-$, $N$=0, $J$=1 & 2.80 & 1150 & 0.72 & 1.51 & 0.94 & \cite{Borin1999}\\
   \botrule
   \end{tabular}}
 \label{table1}
 \end{center}
\end{table}

The structure parameters used for the BBR absorption rate calculations are collected in Table \ref{table1}. We found rates slightly smaller than 1\,s$^{-1}$ (see Table \ref{table2}), suggesting that the lifetime of a molecular sample in a well-defined initial state could be limited by BBR. 
In the near future, molecular interferometry for precision measurements will be performed in traps, which offer the opportunity to keep the molecules for sconds and hence in an interferometry experiment to let the phase being built for much longer times than in a beam experiment. 
Traps are therefore a very promising tool for measuring very small level energy differences with a molecular interferometer, as planned for instance with YbF and PbF.
If seconds are needed to let the phase build up, cycles of absorption-emission are then induced by the BBR and limit the constrast of the interferometer by breaking its coherence.

As a check of such predictions, we also calculated the BBR absorption rate for YbF and PbF using the ionic model and found a good agreement with the calculation using the derivatives of the dipole moment function from refs.\cite{Dolg1992, Liu1998}. 
Moreover, the influence of the inaccuracy on the static polarizability of the ions, which determines the molecular dipole moment, is exemplified with HgF and BaF. 
Indeed, the F$^-$ effective polarizability depends {\it a priori} on the accompanying positive ion. 
In particular, the dipole moment of HgF (2.85\,a.u.) presently obtained using the Hg$^+$ polarizability from ref.\cite{Angstmann2006} (19\,a.u.) is about 70\% larger than the one given in ref.\cite{Dmitriev1992}. 
If we use the ionic model to adjust the dipole moment of HgF to the experimental value, the F$^-$ effective polarizability becomes close (14\,a.u.) to the one of the free ion. 
The derivative of the HgF dipole moment around $R_e$ increases with the F$^-$ effective polarizabilty, increasing the BBR absorption rate only from 0.47\,s$^{-1}$ to 0.86\,s$^{-1}$. 
Note also that Angstmann {\it et al} \cite{Angstmann2006} computed a value of 122\,a.u. for the Ba$^+$ polarizability, significantly higher than in ref.\cite{Torring1984}. 
Introducing this value in the ionic model then yields a F$^-$ effective polarizability of 1.3\,a.u. to recover the experimental dipole moment from \cite{Ernst1986}. We clearly reach here the limits of the validity of the ionic model but this does not significantly change the conclusions of our analysis.

\section{BBR absorption rates for alkali dimers, alkali hydrides and sulfur oxide}
\label{sec:other}

Polar molecules are expected to exhibit long-range dipole-dipole interactions.
One expects to be able to characterize the dipole-dipole
interactions by studying collisions in a cold sample of polar molecules, just like were investigated the details of interactions between cold atoms in the late 90' \cite{Weiner1995}. Cold (half-)collisions could also be studied by controlling the orientation of the molecules with external electric fields. Polar molecules are also promising candidates for quantum computing \cite{Baranov2002,DeMille2002}.

Great experimental efforts are put on the production of ultracold heteronuclear alkali dimers, which possess rather large dipole moments \cite{Aymar2005}. 
They are formed experimentally from trapped ultracold atoms and are therefore obtained at very low temperatures, much lower than in experiments starting from a molecular beam.
The disadvantage of this technique is that it has so far adressed only alkali dimers.  
Up to now densities of ultracold polar alkali dimers are still low, but several group head toward
accumulating and trapping these dimers in order to start investigating molecule-molecule collisions
\cite{Sage2005,Mudrich2004,Mancini2004}.

Many efforts are done nowadays on cold intense beams of alkali hydrides \cite{Tokunaga2007}.
The lightest of them, LiH, could be stopped with a Stark decelerator in order to perform further sympathetic or evaporative cooling in a trap. However, in the samples of trapped molecules obtained up to now \cite{Meerakker2006}, densities are still too low, and temperature still too high to observe cold collisions. 

Due to their large dipole moment, these molecules can undergo vibrational and rotational BBR induced electric dipole transitions within their electronic ground state.
Nevertheless, for all alkali dimers, rotational constants are on the order of a fraction of a
wavenumber, where the spectral energy density of the BBR is very small (see Figure \ref{fig:bbr}).
Only vibrational transitions are therefore of relevance. 
The vibrational matrix elements of the dipole moment were evaluated using the Fourier grid Hamiltonian method \cite{kokoouline1999} taking
as input data electronic ground state potentials and dipole moment functions provided by ref.\cite{Aymar2005}.
We performed the calculations for several alkali dimers (see Table \ref{table2}).
For most of them, the absorption of black body photons cannot be noticeable at room temperature even for extremely long trapping times of several minutes which are reachable if special care is taken to reduce the background gas pressure.
A noticeable exception to this is the case of LiCs. 
The black body limited lifetime of LiCs in its ground state at room temperature is only on the order of one minute, which constitutes a limitation in the investigation of collisions in a trapped cloud of LiCs and further evaporative cooling \cite{Mudrich2004}.

The lowest level in which LiH is foreseen to be efficiently manipulated in an electrostatic
deceleration and trapping process is the $v$=0, $J$=1 rovibrational level of its $X^1\Sigma^+$ electronic ground state.
The pumping rate at room temperature due to vibrational transitions is rather small (0.06\,s$^{-1}$) since the vibrational
constant is quite high ($\approx$\,1400\,cm$^{-1}$).
However, the black body photons are quite efficient at pumping the $v$=0, $J$=1 level to the $v$=0, $J$=2 level at a
rate of 1.26\,s$^{-1}$ and at depumping it down to the $v$=0, $J$=0 level at a rate of 0.163\,s$^{-1}$.
Moreover, the spontaneous emission depletes the $v$=0, $J$=1 level which decays to the $v$=0, $J$=0 level with a rate of 0.036\,s$^{-1}$.
Finally this leads to an overall lifetime of LiH in the $v$=0, $J$=1 level of about 650\,ms at room temperature.
This harms long trapping of LiH in this state and therefore all cooling processes that could follow like evaporative, sympathetic or cavity-assisted cooling.
Even at 77\,K the lifetime increases only to about 3\,s, still limited by the black body assisted rotational transitions.

In order to investigate dipole-dipole interactions, the best alkali hydride is CsH (with a dipole moment of 8.3\,D), for which one can expect strong interactions between
CsH molecules through the dipole-dipole interaction. 
However its dipole moment to mass ratio is not encourageous in the perspective of a Stark deceleration. 
Nevertheless, under the same conditions as for LiH above, we found that the contribution of the vibrational transition to $v$=1 levels amounts to 1.17\,s$^{-1}$,
since the vibrational frequency (891\,cm$^{-1}$) is quite close to the peak of the black body spectrum.
The rotational transitions contribute to the depletion of the $v$=0, $J$=1 level with the following
rates: 0.56\,s$^{-1}$ for $J$=2$\,\leftarrow\,$$J$=1 and 0.071\,s$^{-1}$ for
$J$=0$\,\leftarrow\,$$J$=1, while the spontaneous emission rate is negligibly small (0.0056\,s$^{-1}$). 
In total, at room temperature the $v$=0, $J$=1 level is depleted at a rate of 1.80\,s$^{-1}$. At 77\,K
this rate drops down to 0.16\,s$^{-1}$, essentially given by the $J$=2$\,\leftarrow\,$$J$=1 black body
assisted transition.

Cold chemistry involving free radicals is a blooming field and is expected to bring new insight on
reactive collision and dissociation.
As an experiment aims at producing cold SO radicals and O atoms produced from near-threshold
dissociation of trapped SO$_2$ \cite{Jung2006}, it is worth calculating also the BBR absorption rate of
the SO radical in its ground state. 
The lowest electronic states have been studied and potential energy curves and dipole moments have been
computed \cite{Borin1999}. We estimate that the BBR absorption rate, solely due the first vibrational
transition (1120\,cm$^{-1}$), is $8.3\,10^{-3}$\,s$^{-1}$, in agreement with \cite{Hoekstra2007}. 
This should not affect the trapping time of SO radicals in an electrostatic or in a magnetic trap.

\begin{table}
 \begin{center}
  \tbl{BBR absorption rate at 300\,K for selected heavy diamagnetic polar molecules used in various
precision measurement experiments, for selected alkali dimers and for LiH, CsH and SO. For all calculated rates and lifetimes the initial state is the vibrational ground state ($v$=0). Values for KRb (resp. RbCs) are in agreement with previous calculations of ref. \cite{Kotochigova2003} (resp. \cite{Kotochigova2005}).}
 {\begin{tabular}{l|l|l|l|l}\toprule
   Species & Initial state 
   & $\Gamma_{\text rot}$\,(s$^{-1}$)  & $\Gamma_{\text vib}$\,(s$^{-1}$)
   & Lifetime\,(s)  \\
  \toprule
   YbF  & $X^2\Sigma^+$ & - & 0.26 & 3.8 \\
   BaF  & $X^2\Sigma^+$ & - & 0.60 & 1.7 \\
   HgF  & $X^2\Sigma^+$ & - & 0.47 & 2.1 \\
   PbF  & $X^2\Sigma^+$ & - & 0.29 & 3.4 \\
  \toprule
   LiRb & $X^1\Sigma^+$, $J$=0 & - & 8.0\,10$^{-3}$ & 125\\
   LiCs & $X^1\Sigma^+$, $J$=0 & - & 1.7\,10$^{-2}$ & 59\\
   NaRb & $X^1\Sigma^+$, $J$=0 & - & 7.3\,10$^{-4}$ & 1.4\,10$^3$\\
   NaCs & $X^1\Sigma^+$, $J$=0 & - & 1.7\,10$^{-3}$ & 600\\
   KRb  & $X^1\Sigma^+$, $J$=0 & - & 7.5\,10$^{-6}$ & 1.3\,10$^5$\\
   KCs  & $X^1\Sigma^+$, $J$=0 & - & 8.2\,10$^{-5}$ & 1.2\,10$^4$\\
   RbCs & $X^1\Sigma^+$, $J$=0 & - & 1.5\,10$^{-5}$ & 6.7\,10$^4$\\
   \toprule
   LiH  & $X^1\Sigma^+$, $J$=1 & 1.52 & 6.0\,10$^{-2}$ & 0.63 \\
   CsH  & $X^1\Sigma^+$, $J$=1 & 1.55 & 1.2            & 0.36 \\
   SO   & $X^3\Sigma^-$, $N$=0, $J$=1 & - & 8.3\,10$^{-3}$ & 120 \\
   \botrule
   \end{tabular}}
\label{table2}
 \end{center}
\end{table}

\section{The black body radiation induced shift in molecules}
\label{section:lightshift} 

According to quantum electrodynamics theory, light can not only be absorbed by a massive particle, but it also shifts the level energies of the particle \cite{book:cohen}.
The BBR is known for a long time to be responsible for a shift of the atomic hyperfine splittings \cite{Itanov1982}, which
is of extreme relevance for atomic clocks \cite{Angstmann2006a, Beloy2006, Porsev2006}. 
For atoms in their ground state, all electric dipole allowed transitions require photons with much more energy than available in the BBR spectrum at room temperature. Therefore
the BBR electric field acts on the atom through its static polarizability, exactly like a dc electric field does, and only the amplitude of the black body electric field is then of importance.
As shown in the previous sections, a molecule likely absorbs resonantly a black body photon through an electric dipole transition.
Therefore the BBR does not act on the molecule simply like a dc electric field but the entire BBR spectrum has to be considered. 

The black body is a very incoherent source of photons, with a coherence time of about $h/4 k_B T
\approx 40$\,fs at room temperature \cite{Donges1998}, which is much smaller than the typical
absorption time of a black body photon by the molecule (see rates of Table \ref{table2} and ref.\cite{Hoekstra2007}). 
Therefore, according to ref.\cite{book:cohen}, the BBR-induced frequency shift of the level $i$ reads:

\begin{equation}
\label{BBRlightshift}
\Delta _{i}
=
\frac{4 \pi}{3 \epsilon_0 h c^3}
\sum_\text{$f$ states}
\mathcal{P}
\int d\nu
\frac{\nu^3}{e^{h\nu / k_B T}-1}
\frac{\mu_{fi}^2}{\nu-\nu_{fi}} \,,
\end{equation}
with the same notations as in section \ref{sec:model}, and where the symbol $\mathcal{P} \int$ holds for the Cauchy principal part integral.

Molecular interferometers constitute excellent tools to quantify very small energy differences or measure very accurately transition frequencies. 
First, the experiments carried out on YbF in order to detect the electron EDM ultimately need to measure frequencies down to a precision of tens of $\mu$Hz \cite{Hudson2002}. In that reference, the effect of the BBR is not explicitely discussed. We found that the BBR-induced shift due to rotational electric dipole allowed transitions ($N$=1$\leftarrow$$N$=0) is on the order of tens of mHz, while the contribution to the shift of vibrational transitions is on the order of a few mHz, much larger than the above needed precision. However, a careful analysis of the paper by Hudson {\it et al} \cite{Hudson2002} shows that the ratio of the differences of phases measured under various external field conditions is insensitive to the first order to the induced light shifts. Indeed, each individual phase measured in the YbF interferometer depends on the difference between light shifts induced on two Zeeman sublevels, which is proportional (with an extremely small factor of about $10^{-13}$) to their energy difference. This ensures that a BBR-induced correction to the electron EDM could be at maximum $10^{13}$ times smaller than its actual value, hence totally undetectable. 

The discussion above suggests that a BBR induced light shift could play a significant role if a frequency measurement aimed at a relative accuracy of $10^{-13}$, which is typical of what is needed to investigate possible time variation of fundamental constants. For instance, a very sensitive probe for a possible time variation of the proton-electron mass ratio is the inversion frequency in ammonia. A new experiment is currently being set up, which aims at performing very high resolution spectroscopy in a Ramsey type interferometry experiment in an ammonia molecular fountain \cite{priv:Rick}. An accuracy of $10^{-13}$-$10^{-14}$ is expected on the measurement of the inversion frequency (of about 23.8\,GHz) in the state $J$=1,\,$K$=1. Note that the hyperfine structure of $^{15}$ND$_3$ molecules has been recently measured with an accuracy of a few $10^{-9}$ in spectroscopy experiments using cold decelerated ammonia molecules \cite{veldhoven2004}.
The use of a molecular interferometer is expected to increase further the accuracy by several orders of magnitude.  
Therefore we evaluated the BBR-induced energy shift on both inversion levels involved in this experiment, which might be of importance since the levels considered are of different symmetries, i.e., are coupled to different sets of levels by the electric dipole interaction.

Microwave transitions, which have led to the ammonia maser \cite{Gordon1955}, have been studied for decades in NH$_3$ \cite{Good1947, Costain1951}, as well as its rotation-inversion structure.
We denote the lowest level involved in the inversion transition by $0^-$,\,$J$=1,\,$K$=1, the upper one being $0^+$,\,$J$=1,\,$K$=1, with the opposite parity.
Both levels are coupled by electric dipole allowed transitions to all other rovibrational levels.
Let us examine first the rotational radiative coupling: each of these levels is coupled to an inversion level of the $J$=2, $K$=1 rotation manifold {\it via} an electric dipole allowed transition. 
Because of the inversion-rotation structure (see for instance \cite{book:SchawlowTownes}), these two allowed transitions, around 39.7\,cm$^{-1}$, differ in frequency by the sum of the inversion splittings of the $J$=1, $K$=1 and $J$=2, $K$=1 states, i.e., by about 46.8\,GHz. 
This implies that the levels $0^-$,\,$J$=1,\,$K$=1 and $0^+$,\,$J$=1,\,$K$=1 experience different BBR-induced shifts, given by eq.(\ref{BBRlightshift}).
Both BBR-induced shifts, on the order of 80\,mHz at room temperature, differ by about 5\,mHz, which represents 2\,10$^{-13}$ times the inversion frequency between the $0^-$,\,$J$=1,\,$K$=1 and $0^+$,\,$J$=1,\,$K$=1 levels. 
Although not negligible in the absorption rate of BBR photons, the coupling to higher vibrational states {\it via} BBR photons only gives rise to light shifts lower than 10$^{-14}$ times the inversion frequency (see for instance \cite{Yurchenko2005a,Yurchenko2005b} for electric dipole moment matrix elements).
Note that the dipole moment matrix elements we used give an absorption rate compatible with the one calculated in ref.\cite{Hoekstra2007}.

In conclusion, at room temperature the measured inversion frequency between the $0^-$,\,$J$=1,\,$K$=1 and $0^+$,\,$J$=1,\,$K$=1 levels is lowered by about 2\,10$^{-13}$ by BBR-induced energy shifts.
We have also calculated that a room temperature variation of 5$^{\circ}$C changes relatively the measured frequency by about $10^{-14}$. 
We did the same type of calculation on $^{15}$ND$_3$ and we draw the same conclusion, i.e., that the inversion frequency is lowered by about 2\,10$^{-13}$ by BBR-induced shifts.
This should be considered carefully in a high-precision interferometry experiment aiming at measuring the inversion frequency of ammonia at an accuracy of $10^{-13}$-$10^{-14}$ and more generally in high-precision frequency measurements involving molecules.

\section{Conclusion}
We have performed calculations to evaluate the influence of the black body radiation on molecular systems in use in, or of relevance for high-precision measurements.
Both absorption rates and induced energy shifts have been estimated. 
The BBR absorption could disturb the coherence of molecular interferometers if they are achieved in traps with heavy radicals like YbF, PbF or HgF, which hold promise in the search for the electron EDM.
Alkali dimers should not suffer from the BBR absorption, except LiCs that survives in its ground state during one minute only, which might hamper further evaporative or sympathetic cooling.
Much more critical is the case of LiH, which is predicted to be pumped out of the prepared initial state $v$=0, $J$=1 in less than one second at room temperature and only in 3\,s at liquid nitrogen temperature.
Finally, we have shown that the BBR-induced shift on molecular level energies should be considered carefully in high-precision frequency measurements, like in the case of a molecular fountain of ammonia, which is currently being built to investigate the time variation of the proton-electron mass ratio. 

\section{Acknowledgements}
Useful discussions with Christian Jungen are gratefully acknowledged.
We are indebted to Mireille Aymar for providing dipole moment functions of alkali hydrides.
This work has been supported by the "Institut Francilien de Recherche sur les Atomes Froids" (IFRAF) and by the "Agence National de la Recherche" (ANR grant NT05-2 41884).


\end{document}